\documentclass[referee]{aa} 

\usepackage{graphicx,hyperref}
\usepackage{txfonts}
\usepackage{xcolor}

\newcommand{\mJyb}{mJy~beam$^{-1}$}
\newcommand{\uJyb}{$\mu$Jy~beam$^{-1}$}
\newcommand{\tar}{J1430+2303}

\usepackage{amstext}

\begin{document} 

\title{ VLBI imaging of the pre-coalescence SMBHB candidate SDSS~J143016.05+230344.4}
\titlerunning{VLBI imaging of J1430+2303}
   \author{Tao An
          \inst{1,2,3}
          \and
         Yingkang Zhang\inst{1}
         \and 
         Ailing Wang\inst{1,3}
         \and 
         Xinwen Shu\inst{4}
         \and
         Huan Yang\inst{5,6}
         \and
         Ning Jiang\inst{7,8}
         \and
         Liming Dou\inst{9}
         \and
         Zhen Pan\inst{5}
         \and
         Tinggui Wang\inst{7,8}
         \and
         Zhenya Zheng\inst{1}
          }

   \institute{Shanghai Astronomical Observatory, Chinese Academy of Sciences, Nandan Road 80, Shanghai 200030, China\\
              \email{antao@shao.ac.cn}
         \and
             Key Laboratory of Cognitive Radio and Information Processing, Guilin University of Electronic Technology, 541004 Guilin
         \and
             College of Astronomy and Space Sciences, University of Chinese Academy of Sciences, 19A Yuquanlu, Beijing 100049, China
      \and   Department of Physics, Anhui Normal University, Wuhu, Anhui, 241002, China       
        \and
            Perimeter Institute for Theoretical Physics, Waterloo, ON N2L2Y5, Canada
        \and
            University of Guelph, Department of Physics, Guelph, ON N1G2W1, Canada
        \and    
             CAS Key Laboratory for Research in Galaxies and Cosmology, Department of Astronomy, University of Science and Technology of China, Hefei, Anhui 230026, China
        \and
            School of Astronomy and Space Science, University of Science and Technology of China, Hefei 230026, China
        \and
            Department of Astronomy, Guangzhou University, Guangzhou 510006, China
}

   \date{Received 4/20/2022; accepted 5/3/2022}

 
  \abstract
   {Recently, SDSS J143016.05+230344.4 (J1430+2303) was reported to be a supermassive black hole binary (SMBHB) in the final coalescence phase. It is probably the first SMBHB coalescence event observable in human history. Radio observations of J1430+2303 before and after coalescence will provide a unique diagnosis of the energetics and environment of the SMBHB.}
   {We explore the radio emission from the galactic nucleus region that is closely related to the current X-ray and optical activities and helps to understand the state of black hole accretion and outflow before coalescence.}
   {Very long baseline interferometry (VLBI) imaging is the only method that offers milli-arcsecond-level high resolution that can exclude the contamination by diffuse emission on galactic scales. 
   We observed \tar\ with the European VLBI Network at 1.7 GHz and with the Very Long Baseline Array at 1.6 and 4.9 GHz in late February and early March 2022.}
   {A compact component is detected in all three VLBI images. It has a brightness temperature of $>10^8$~K, an unresolved morphology with a size $<0.8$ pc, and a flat radio spectrum. These observational features are inconsistent with large opening-angle outflows or winds, but indicate that this compact component might 
   be
   a jet or a corona. Nearly 60\% of the emission is resolved by VLBI and may come from remnant lobes of previous radio activities, the outer layers of a structured jet, or shocks formed by the disc winds in the narrow line region.}
   {Current VLBI images do not yet show signs of radio outbursts. Our observations provide pre-coalescence radio data that are an important reference for future comparative studies with the post-merger. In particular, further 
   resolving  
   the jet will pave the way for probing the dynamical features associated with inspiralling binary black holes.}


   \maketitle
%

\section{Introduction} \label{sec:intro}

Most massive galaxies in the Universe have at least one supermassive black hole (SMBH) at their centres \citep[e.g.][]{2013ARA&A..51..511K}. During the hierarchical formation of galaxies, SMBHs are expected to migrate towards the galactic centre through star scattering and/or gas friction, eventually forming an SMBH binary (SMBHB) on parsec scales \citep{1980Natur.287..307B,2003ApJ...582..559V}. Observations of SMBHBs will provide  astronomers with information about the evolutionary path and history of galaxies towards their present morphology, that is, the shape and the size. Most previous studies of SHBHB evolution have assumed circular orbits for simplicity, and/or expected that the gas friction or star scattering may dampen out the orbit eccentricity. However, if the orbits of SMBHBs are highly eccentric \citep{2005ApJ...634..921A,2006ApJ...642L..21B,2011ApJ...729...13C}, the strong-gravity regime extends to larger orbital separation and earlier evolution age, such that SMBHBs may rapidly evolve to coalescence due to gravitational wave (GW) emission. 

With modern observational facilities,  even using the highest resolution of the very long baseline interferometry (VLBI) technique, direct identification of close SMBHBs with a separation $\lesssim$1 pc from radio imaging is very challenging \citep{2018RaSc...53.1211A}. 
In the past decade, 
the rapid development of
time-domain observations 
has
promoted a non-imaging technique for searching for SMBHBs. This method targets periodic variations in the optical  luminosity or emission line intensity of active galactic nuclei (AGNs) over sufficiently long timescales. If the underlying source is indeed an SMBHB, the periodic variability on a timescale comparable to the SMBHB orbital period may originate from accretion rate fluctuations \citep{2008ApJ...672...83M,2012ApJ...755...51N,2015Natur.518...74G} or relativistic Doppler modulation \citep{2015Natur.525..351D}. 
The famous SMBHB candidate OJ~287  \citep{1988ApJ...325..628S,1996A&A...305L..17S} provides a nice demonstration of this technique. 
On the other hand, a number of SMBHB candidates with inferred orbital sizes of 0.001--0.01 pc have been identified  using this time-domain method  \citep{2015Natur.518...74G,2014ApJ...789..140L,2016ApJ...827...56Z,2020MNRAS.499.2245C}, although it is worth mentioning that the interpretations of their optical variability periods remain highly controversial  \citep{2015Natur.525..351D,2016MNRAS.461.3145V,Zhu2020}. A major concern is that the red noise in the AGN light curve may also produce apparently periodic signals. In order to rule out the cause of AGN red noise, Bayesian model selection is necessary to address the statistical significance of the observed signals. 
Furthermore, all known SMBHB candidates so far, including OJ~287, have relative large orbital separations. Assuming that only GW radiation is at play, their expected merger timescales exceed the Hubble time. It is therefore possible that they belong to a class of massive black hole binary sources different from those observable by space-borne gravitational wave observatories, such as \textit{LISA}, \textit{Tianqin,} and \textit{Taiji} \citep{2021NatAs...5..881G}.

Recently, \citet{2022arXiv220111633J} reported the discovery of the SMBHB candidate SDSS~J143016.05+230344.4 (also called AT2019cuk; hereafter J1430+2303). It is located in a Seyfert type I galaxy at a redshift of 0.08105 \citep{2015ApJS..219....1O}. The  optical light curves obtained from Zwicky Transient Facitity (ZTF)  display a quasi-periodic pattern of variability (Figure 1 therein), and more intriguingly, the periodicity exhibits a trend of rapid decay in both period and amplitude. In the past three years,  the period has decreased from $\text{an approximately}$ one-year cycle at the first peak in 2019 July to a fourth cycle with a period  $ \text{of about three}$ months in 2021 June. No such chirping flares with decaying periods have been observed  before. They are incompatible with known models of disc  oscillations or disc  instability \citep{2019Natur.573..381M,2008Natur.455..369G}.  The analysis in  \citet{2022arXiv220111633J} shows that the timings of the flares are consistent with the theoretical expectation of an SMBHB near its final coalescence. A Bayesian model selection analysis also disfavours AGN red noise as the origin of the chirping flares. Together with the X-ray and the hydrogen emission line measurements, the SMBHB scenario becomes the most plausible interpretation of the various models considered. 

If the SMBHB model is confirmed, \tar\ would be the first SMBH coalescence ever observed. 
In the SMBHB scenario, the orbit is expected to be inclined (with respect to the accretion flow) and highly eccentric. Moving SMBHs collide with the accretion flow close to the pericentre and 
expel
a certain amount of gas, forming  expanding plasma balls that shine in the optical band as  flares. The X-ray emission from this system is more complicated as it is susceptible to the variation of accretion conditions in the inner accretion disc.
On the other hand, because the orbit is highly eccentric, a significant amount of the GWs is emitted near the pericentre passage, which causes rapid orbital decay, as evidenced by the timing of the flares. 
The orbital evolution model predicts that coalescence will occur within  three years \citep{2022arXiv220111633J}. During the SMBHB inspiral stage, especially close to the final merger, a large amount of electromagnetic radiation is expected to be released in the X-ray to optical bands \citep{2005ApJ...622L..93M}. 
Following the black hole coalescence, a relativistic jet may be produced in the pre-merger wind bubble with non-thermal radiation, forming a slowly fading transient that yields an opportunity for monitoring at electromagnetic wavelengths \citep{2021ApJ...911L..15Y}. 
Therefore,  observations across the entire electromagnetic spectrum before and after the coalescence are essential for a comprehensive understanding of this rare astrophysical phenomenon. 

In particular, observations of the radio emission from \tar\ will provide a unique diagnostic of the energetics and environment of the SMBHB coalescence. Shocks can be generated during the disc crossings, which subsequently affect the AGN corona and produce synchrotron radiation in the radio band; ultra high velocity outflows observed in the optical emission lines and X-ray continuum may also produce shocks on larger scales beyond the broad line region (BLR). In addition, after the binary merger, as the accretion flow gradually settles on the coalesced black hole, a transient jet may be produced in a way similar to in tidal disruption events. The X-ray and optical monitoring campaigns
are ongoing, and simultaneous radio observations will add valuable data for studying the accretion physics and evolution of the SMBH(s). 

J1430+2303 was previously observed by the Very Large Array (VLA) FIRST Survey (Radio Images of the Sky at Twenty-Centimeters) at 1.4 GHz \citep{1995ApJ...450..559B}, and by the Very Large Array Sky Survey (VLASS) at 3 GHz \citep{2020PASP..132c5001L}. A weak source of $\sim1.0 \pm 0.2$ mJy is seen at the position of \tar\ in the FIRST image, observed on 1995 December 18. The source is detected with a flux density of $\sim0.6 \pm 0.1$ mJy in the 3 GHz VLASS image observed on 2017 November 17, but 
is not unambiguously detected
on 2020 July 16. In the Rapid ASKAP Continuum Survey (RACS) at 888 MHz \citep{2020PASA...37...48M}, a compact component of $\sim0.9 \pm 0.2$ mJy is found at the position of the FIRST source on 2020 May 1. 
A recent 
enhanced Multi Element Remotely Linked Interferometer Network (e-MERLIN
observation made on 2022 March 28 reported a compact source with a size of 85 mas $\times$ 34 mas and a flux density of $0.73 \pm 0.09$ mJy (ATel\#15306).
These data suggest that the flux density of \tar\ between 0.9 and 1.5 GHz is in the range of 0.7--1.0 mJy.
Moreover, these archival data indicate that variability is unlikely to be the cause of the VLASS non-detection in 2020, while the quality of the 2020 data calibration is worthy of attention.

In order to measure the radio spectrum of \tar\ and detect possible radio signals from the SMBHB coalescence, we applied for the director's discretionary time (DDT) of the Australia Telescope Compact Array (ATCA), the Karl G. Jansky Very Large Array (JVLA), and the upgraded Giant Metrewave Radio Telescope (uGMRT), covering frequencies from 0.7 GHz to 22 GHz. 
ATCA, VLA, and uGMRT have resolutions on the order of arcseconds and are able to obtain the radio emission from the whole galaxy, but are unable to probe the state of radio emission from the galactic nucleus. High-resolution VLBI imaging is the only observational method with milliarcsecond-scale (corresponding to parsec-scale physical size) resolution that can confirm the presence of a compact radio core from its compactness and high brightness temperature, and possibly  detect jet knots. Detection of the jet will pave the way for future monitoring of dynamical signatures associated with the inspiral binary black holes, such as the variation in jet position angle. Therefore, we applied for observations of \tar\ with the European VLBI Network (EVN) and the US Very Long Baseline Array (VLBA). The main goal of this round of pilot VLBI observations is to use high-resolution images to verify the presence of compact radio source(s) in the galactic nucleus and to obtain their precise positions so as to provide a priori information for future VLBI monitoring programs.

This paper focuses on our VLBI observations. The multi-frequency radio spectrum based on the ATCA, JVLA, and uGMRT is presented in a separate paper (Shu et al. in prep.). Details of the EVN and VLBA observations are given in Section \ref{sec:obs}, the results are presented in Section~\ref{sec:result}, and Section~\ref{sec:conc} summarises our study. In this paper, we assume a standard flat $\Lambda$ cold dark matter cosmology with $\Omega_{\rm m} = 0.27$, $\Omega_{\rm \Lambda} = 0.73$, $H_0 = 70$~km\,s$^{-1}$ Mpc$^{-1}$, and the 1 mas angular scale corresponds to a projected physical size of 1.53 pc at the redshift of \tar. 

\section{Observations and data processing} \label{sec:obs}

\subsection{VLBA observations} \label{sec:vlbaobs}

The initial VLBA observations, approximately two weeks following the application of the DDT, were completed on March 2 UT 08:17 -- March 3 UT 0:44 at L band (central frequency of 1.57 GHz) and on March 3 from UT 08:49 to UT 13:19 at C band (central frequency of 4.90 GHz) (observation code: BA154). All ten VLBA antennas participated in the observations. Each session lasted for 4.5 hours (Table~\ref{tab:obs}). The total flux density of \tar\ is $\lesssim$1 mJy at 1--3 GHz from the archive VLA observations (see Section \ref{sec:intro}), therefore the phase-reference mode was adopted. 
The pointing centre of the \tar\
was at RA = $14^\mathrm{h} 30^\mathrm{m} 16.056^\mathrm{s}$, Dec = $+23^{\circ} 03^{\prime} 44.45^{\prime\prime}$  (J2000), 
which corresponds to the SDSS peak position and is close to the VLA FIRST peak. 
The positional accuracy of the SDSS and FIRST is about $0.2\arcsec$, which is practicable for VLBI observations in the absence of milliarcsecond-resolution images. The bright source J1427+2348 (200 mJy at 5 GHz, about $1\degr$ from the target) was used as the phase-reference calibrator. The coordinate of J1427+2348 is RA=14:27:00.3918, Dec=+23:48:00.038 with a position error of only 0.1 mas, obtained from the Astrogeo database \footnote{\url{http://astrogeo.org} maintained by L. Petrov.}. The cycle of the phase-reference observations is 6 min, of which 4.5 min were used for the target source and 1.5 min for the calibrator source. The data recording rate was 2~Gbps, that is, using 256~MHz bandwidth (divided into 16 channels, 16~MHz bandwidth per channel), one-bit sampling, and dual polarisations. This configuration enables a field of view of $1.4\arcsec$ centred on the pointing centre, which is less affected by the bandwidth smearing effect.

The raw observational data recorded by the VLBA telescopes were transferred to the DiFX correlator \citep{2011PASP..123..275D} in Socorro, USA, for correlation processing. The integration time of the correlator is 2 s, which can reduce the time smearing effect. After this,  the correlated data were sent to the China SKA Regional Centre \citep{2019NatAs...3.1030A} and were further processed using a VLBI data processing pipeline
for the calibration of the visibility data and imaging. This pipeline is  written in Python and executed in script. It  uses Parseltongue as an interface to call programs from the Astronomical Image Processing System (\textsc{aips}) software package developed by the National Radio Astronomy Observatory (NRAO) \citep{2003ASSL..285..109G} to perform data calibration. The data in two bands were processed separately following the same procedure. 

After loading the raw correlation data, the visibility amplitudes were calibrated using the antenna gain curves and the system temperatures measured at each telescope. We then checked the data quality and removed some obviously odd data points caused by observational failures and radio frequency interference. Throughout the data quality check, we used the Pie Town (PT) antenna as the reference antenna. Considering the weather information in the auxiliary tables, we corrected the data for ionospheric dispersion delay using the total electron content map obtained from the Global Navigation Satellite System (GNSS) data, and we also calibrated the atmospheric opacity effect on the visibility amplitude. For radio telescopes of azimuth-elevation mounted structures, the phase variations due to the temporal variation of the source parallactic angles were corrected as well. A fringe fitting was then performed on the bright fringe-searching source 3C345 arranged in the experiment, and the gain solutions we obtained were used to correct for instrumental delays and phase errors between different sub-bands,with the solutions  applied to all the data. At the next step, we ran the global fringe fitting using the \texttt{FRING} task \citep{1983AJ.....88..688S} on the phase-reference calibrator J1427+2348, and the resulting gain solutions were interpolated and applied to all the data. Finally, the bandpass function of each antenna was solved by using the \texttt{BPASS} task and applied to calibrate the amplitude of all visibility data.

\begin{figure*}
\centering
        \includegraphics[width=0.3\textwidth]{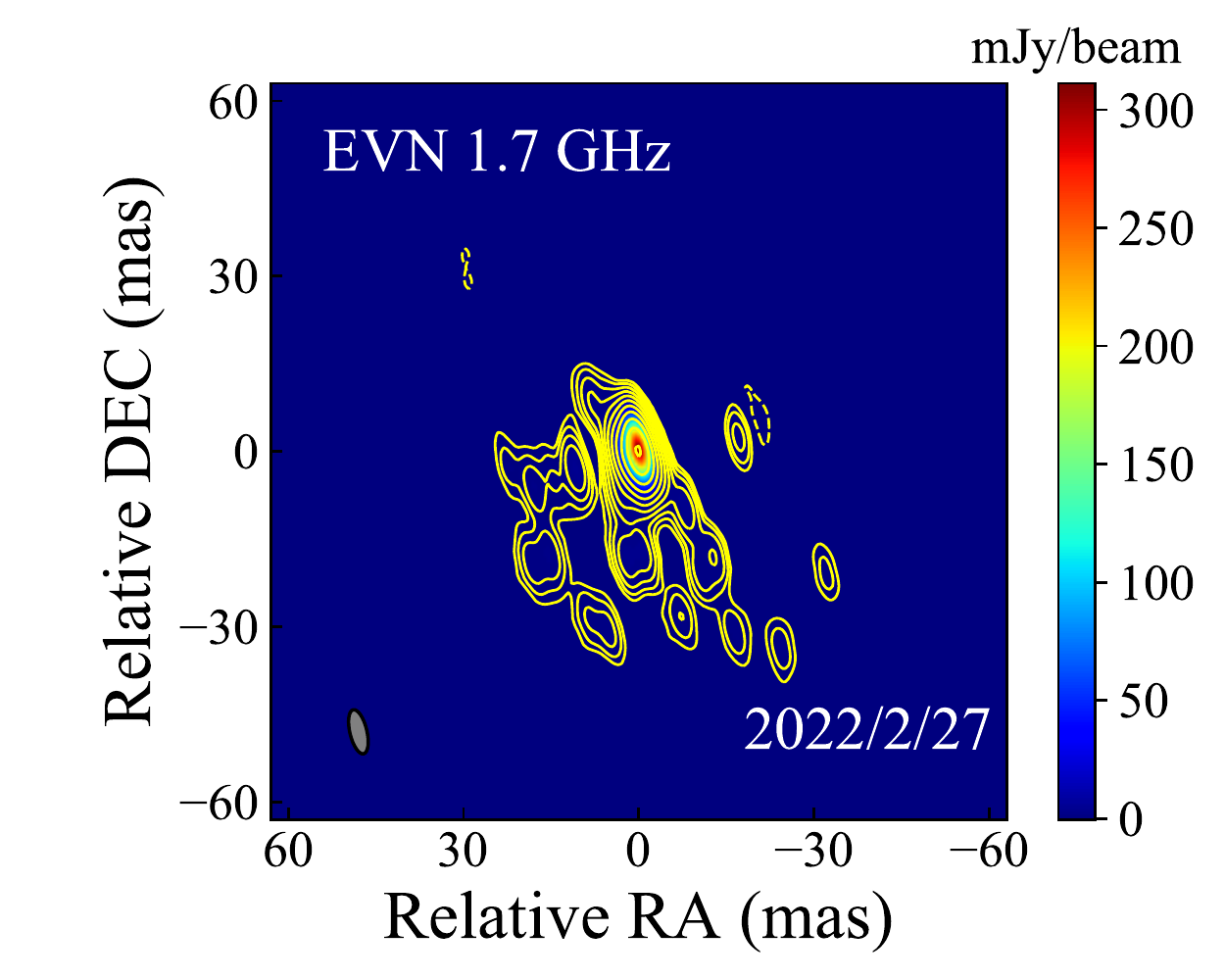}
        \includegraphics[width=0.3\textwidth]{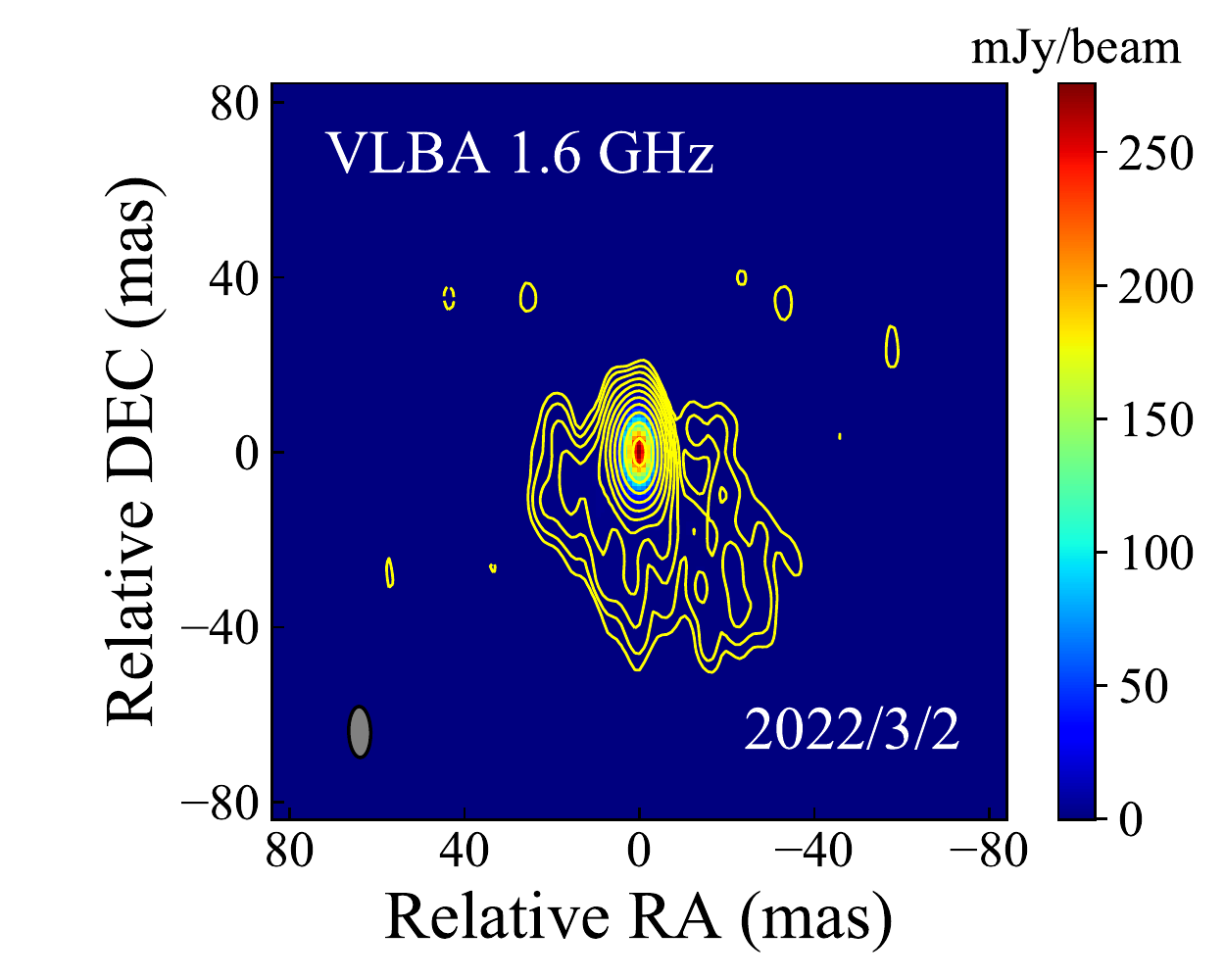}
        \includegraphics[width=0.3\textwidth]{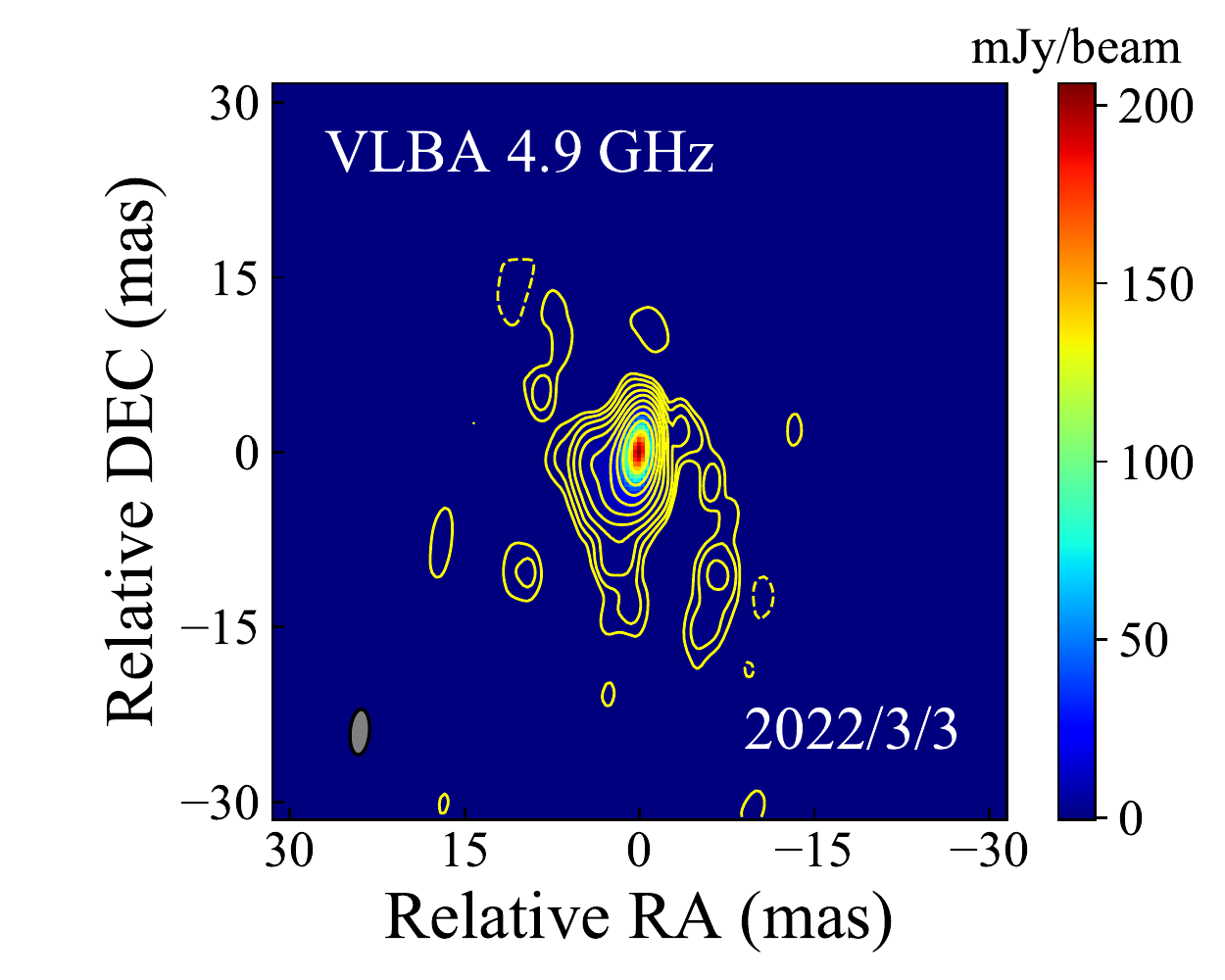}
    \caption{Natural-weighted images of the phase-reference calibrator J1427+2348. In each image, the bottom left ellipse is the shape of the restoring beam. The colour bar denotes the brightness.  \textit{Left}: EVN image at 1.66 GHz. The rms noise is 0.05 \mJyb, and the beam size is $7.7 \times 2.9$ mas$^2$ at a position angle of 13.4\degr; \textit{Middle}: VLBA image at 1.57 GHz. The rms noise is 0.05 \mJyb, and the beam size is $11.7 \times 5.5$ mas$^2$ at a position angle of 1.6\degr; \textit{Right}: VLBA image at 4.9 GHz. The rms noise is 0.04 \mJyb, and the beam size is $3.9 \times 1.7$ mas$^2$ at a position angle of $-$1.3\degr. The lowest contours represent five times of the background rms noise, and the contour levels increase by a factor of 2.}
    \label{fig:cal}
\end{figure*}

With  the  calibration procedures completed, the data of the phase-reference calibrator J1427+2348 were first exported as external single-source FITS files, averaged for each sub-band (128-MHz each) and 2-second sampling time to reduce the amount of data. The J1427+2348 data were then loaded into the \textsc{Difmap} software package \citep{1994BAAS...26..987S} for self-calibration and imaging. After several iterations of hybrid mapping \citep{1974A&AS...15..417H,1984ARA&A..22...97P}, a high-quality image of the calibrator source was obtained.
We then estimated the antenna-based gain correction factors, which primarily reflect the uncertainty in the amplitude of the visibility data associated with the antenna gain curves and the system temperatures. 
We found that the gain correction factor ($g_n$) for most antennas deviated from the nominal value (=1.0) by no more than $\pm10\%$, with an average gain correction factor of $\bar{g} = 0.05$ with a standard deviation of 0.12 at 1.6 GHz, $\bar{g} = 0.02$ with a standard deviation of 0.07 at 4.9 GHz. Therefore, we adopted a mean amplitude calibration uncertainty of 5\% for the VLBA visibility data. 
We applied the correction factors to the visibility amplitudes of all calibrators and the target source in \textsc{aips}. The calibrator model components generated in \textsc{Difmap} were imported into \textsc{aips} as input models for fringe fitting of the calibrator source data. This operation took into account the calibrator structure to improve the phase solutions because it is very important to improve the image quality of the target source when the calibrator is not very compact (see Figure~\ref{fig:cal}). Finally, the fringe-fitting solutions obtained for the phase-reference calibrator were interpolated to the target source data.  We should note that J1427+2348 essentially presents a compact structure at 1.7 GHz, but exhibits a resolved structure at $>5$ GHz: a core-jet structure extending to the south-east, but with a diffuse morphology in the jet beyond 2 mas. Simultaneous multi-frequency observations made by RATAN-600 show a steep radio spectrum above 5 GHz and a turnover below 5 GHz \citep{2012A&A...544A..25M}, indicating an absorption of the compact core-jet structure at low frequencies. Therefore the calibration of the jet structure at 5 GHz is necessary, but not very relevant at 1.6 GHz. With the above calibration procedures, all calibrations were completed. Finally, after applying all calibration solutions to the target source, its visibility data were exported from \textsc{aips} and imported in \textsc{Difmap} for imaging.

\subsection{EVN observation} \label{sec:evnobs}

The EVN observations were carried out from 
2022 February 27 UT 21:30 to February 28 UT 3:00 at a central frequency of 1.66~GHz. e-VLBI mode was used in order to obtain observational results quickly \citep{2004evn..conf..257S}. Fourteen telescopes participated in this observation (see Table~\ref{tab:obs}). During the experiment, data from the radio telescopes were transferred in real time through a high-speed optical fiber network to the SFXC software correlator \citep{2015ExA....39..259K} for processing at the Joint Institute for VLBI European Research Infrastructure Consortium (JIVE) in Dwingeloo, the Netherlands. Data from some individual telescopes were first temporarily stored at the station due to network problems and transferred to JIVE via the internet within two weeks after the end of the observations. The observations were performed in phase-reference mode by rapidly switching the telescopes between the target and a nearby bright calibrator (J1427+2348). The phase errors in the visibility data caused by the atmosphere can be solved by observing the bright phase-reference calibrator and applying the solutions to the target source. The phase-reference cycle was 5 minutes (4 minutes on the target and 1 minute on the calibrator) and repeated 53 times during a total of 5.5 hours observing time to obtain a good (u,v) coverage. 

The SDSS coordinate (J2000) of \tar\ was used for the pointing position of the EVN telescopes. While waiting for JIVE to perform the correlation process, our VLBA observational data were released on March 17, and they were subsequently processed to yield a successful detection of the compact component for \tar. The position of the compact source deviates from the pointing position by about 1~arcsec. In order to obtain the best possible phase coherence and reduce the bandwidth smearing effect, we requested the JIVE  to re-correlate the data at the VLBA peak position. 

The correlated EVN data were transferred to the China SKA Regional Centre and processed  following the same pipeline  described in Section \ref{sec:vlbaobs}. The EVN and VLBA observations were made within 3 days of each other, and  the same phase-reference source and the same cycle was used to facilitate comparison of the two 1.66-GHz observations. Figure~\ref{fig:cal} shows the images of the calibrator J1427+2348, with a complex jet morphology. The brightest central component is the radio core, with a peak flux density of about 250 \mJyb\ at 1.6 GHz and $\sim$200 \mJyb\ at 4.9 GHz. 
The L-band (1.6 GHz) images from the EVN and VLBA appear to be very similar. The core is surrounded by diffuse emission to the east, south, and west. The integrated flux densities of the core derived from two observations agree well, suggesting that the uncertainty of the visibility amplitude calibration is within a few percent. A subtle difference is that the EVN has a higher resolution and is able to reveal a more detailed clump-like structure.
The 4.9-GHz image shows only the inner 8 mas core-jet structure, and the extended emission on a larger scale is resolved out or becomes too weak to be detected at 4.9 GHz because its spectrum is steep.

The amplitude self-calibration from the calibrator J1427+2348 yielded different antenna gain correction factors for individual antennas, in the range of 0.8--2.1. Compared with the VLBA telescopes, which have the same size and the same equipment, the performance of the heterogeneous EVN telescopes shows a relatively larger difference, as is also manifested in the amplitude measurement errors of different telescopes. We calibrated the EVN visibility data in \textsc{aips} using a median correction factor for each telescope. Moreover, we corrected the phase errors caused by the structure of the calibrator itself using the \texttt{CLEAN} model of the calibration source.

\subsection{Model fitting and error estimate of fitted parameters}

Because the target source is very weak, self-calibration is technically difficult. We performed only a few runs of \texttt{CLEAN} and used natural weighting for deconvolution imaging. A circular Gaussian brightness distribution model was used to fit to the visibility data \citep{1995ASPC...82..267P} to quantitatively describe the size and flux density (Table~\ref{tab:img}). However, we find that after several tens of fitting iterations, the circular Gaussian often degenerates into a point source, that is, it has a very small size close to 0. This situation indicates that the radio source has a very compact emission structure and that the resolution of the VLBI image is not sufficient to resolve it. In this case, we cannot measure its size exactly, but  only estimate an upper limit value based on the array configuration and the sensitivity of the image \citep{2005astro.ph..3225L}. We used a method that takes both statistical and measurement errors into account to estimate the errors of the observational parameters. The error in the flux density is the root of the sum of the squares of the image noise and the visibility amplitude calibration error (see Section~\ref{sec:obs}). The astrometric error of the absolute position of \tar\ is mainly sourced from the position error of the calibrator, which is about 0.1 mas. 

\section{Results}\label{sec:result}

\tar\ may have multiple sources of radio emission: star formation activity, a weak jet, accretion-disc-driven winds or outflows, and an accretion disc corona \citep{2019NatAs...3..387P}. The first source of emission (star formation) is on the galactic scale. In principle, the jet is able to extend from near the central engine up to kiloparsec scales, but most of the radio-quiet AGN jets are not powerful enough to break through the confinement of the host galaxy environment ($\lesssim 1$ kpc) \citep{2012ApJ...760...77A}. The outflow mainly affects the broad line region (BLR) and narrow line region (NLR) \citep{2007ApJ...668L.103B}. The size of the corona is very small, but may extend to a fraction of 1 pc in some cases \citep{2008MNRAS.390..847L}. These sources of radio emission are mixed together, but they can be distinguished by morphology, size, brightness temperature, and spectral index.

\subsection{Nature of the parsec-scale compact component}

\begin{figure*}
\centering
        \includegraphics[width=0.3\textwidth]{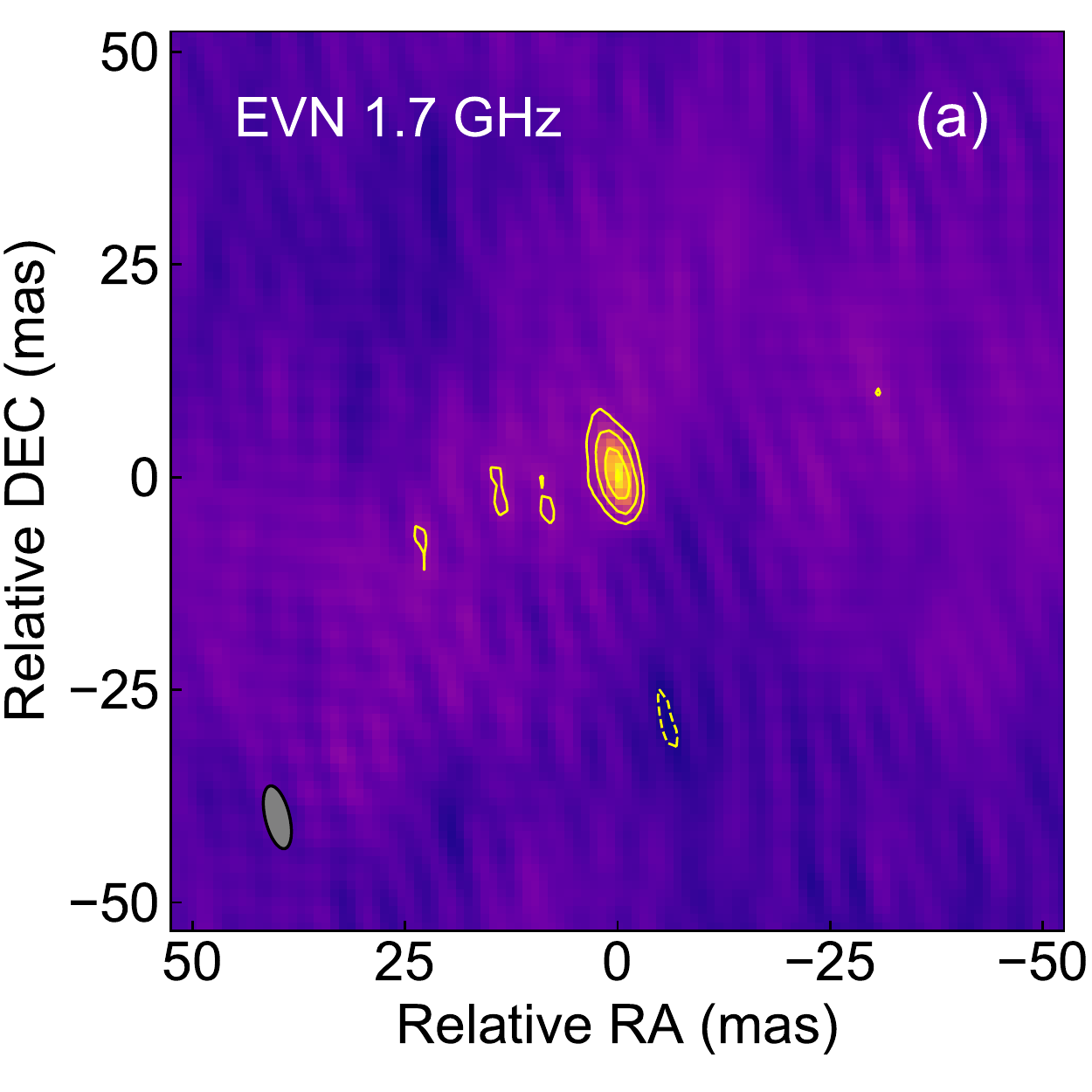}
        \includegraphics[width=0.3\textwidth]{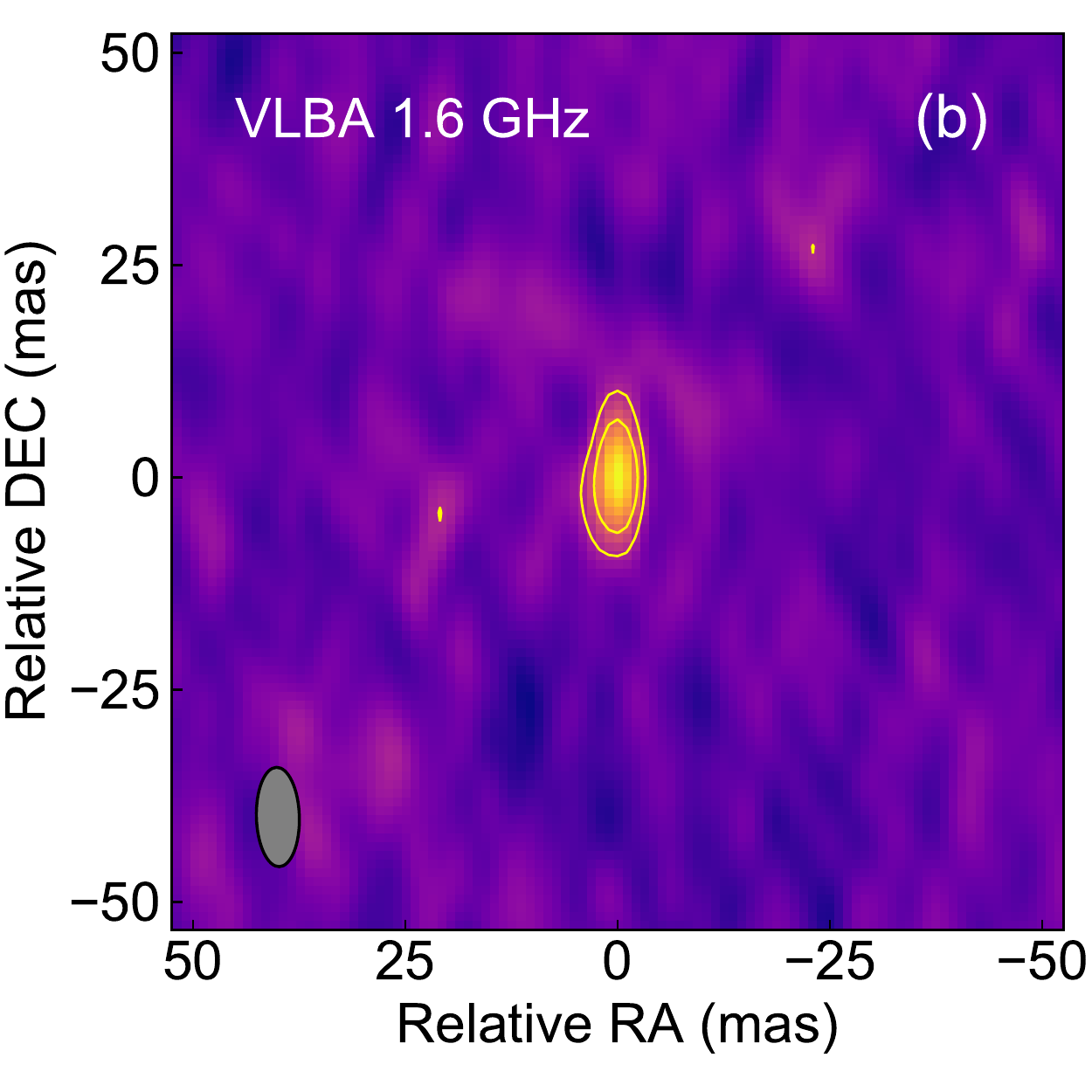}
        \includegraphics[width=0.3\textwidth]{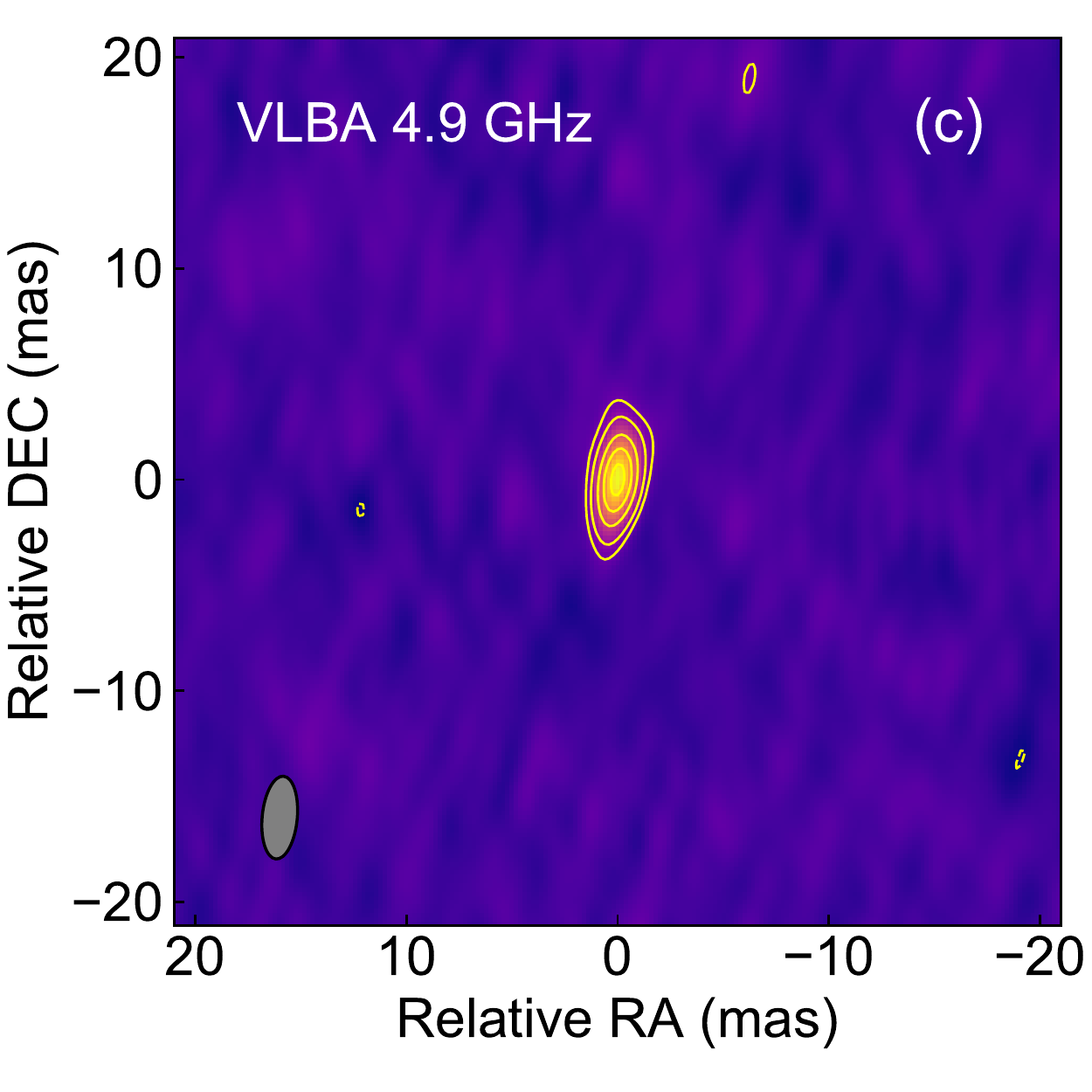}
    \caption{Natural-weighted images of three VLBI observations of J1430$+$2303. \textit{Left}: EVN image at 1.66 GHz. \textit{Middle}: VLBA image at 1.57 GHz. \textit{Right}: VLBA image at 4.9 GHz. In each image, the bottom left ellipse is the shape of the restoring beam. The contour levels are 3$\sigma$ $\times$ ($-$1, 1, 2, 4, 6, 8, 10, 20, and 40).  The image parameters are listed in Table~\ref{tab:img}.}
    \label{fig:map}
\end{figure*}

Figure~\ref{fig:map} shows the images obtained from the EVN and VLBA observations. All images show a compact unresolved component. The difference between the peak positions (Table~\ref{tab:img}) in the 1.57 GHz VLBA and 1.66 GHz EVN images does not exceed 0.3 mas (three times the astrometric error; see Section \ref{sec:obs}). The difference between the peaks in the 4.9 GHz VLBA image and the 1.57 GHz VLBA image is slightly larger. In addition to the astrometric error, the differential optical depth at the two frequencies additionally contributes to the positional error.
In the three figures, the noise is uniformly distributed, with average noise levels of 0.015 \mJyb\ (EVN 1.66 GHz), 0.018\mJyb\ (VLBA 1.57 GHz), and 0.008 \mJyb\ (VLBA 4.9 GHz).
EVN has more antennas and contains large-diameter telescopes such as the Efflesberg 100m and Tianma 65m telescopes, so that the sensitivity of the EVN is slightly higher than that of the VLBA for the same frequency and observation configuration.
The flux density of the EVN component is higher than that of the VLBA component at 1.6 GHz, and the difference is about four times the flux density uncertainty. The EVN and VLBA observations are very close in frequency and observation time. Rapid variability is very unlikely for such a radio-quiet AGN. 
This difference probably results from the observation and data calibration. That is, the correlation of the EVN data used the precise position obtained from the VLBA 4.9 GHz image (see Section \ref{sec:evnobs}), so that the coherence loss in the VLBA data may be the main reason that the flux density is lower than in the EVN data. 

We have used the VLBI observational results to calculate the brightness temperature $T_{\rm B}$ of the radio component \citep{1982ApJ...252..102C}, and the results are given in Table~\ref{tab:img}. As discussed above, the parsec-scale structure of this source is very compact and not resolved in our VLBI images, therefore we only constrained the upper limit of the source size from which the lower limit of $T_{\rm B}$ is obtained. The 4.95 GHz VLBA image yields an upper limit for the component size of 0.5 mas (corresponding to a projected size of $\sim$0.77 pc). The EVN data give the highest $T_{\rm B}$ limit, which is $>1.4 \times 10^8$~K. If a radio source is dominated by thermal emission, the maximum brightness temperature of a normal galaxy is $\lesssim10^5$~K \citep{1992ARA&A..30..575C}. The derived brightness temperature of $\gtrsim 10^8$~K for \tar\ significantly exceeds this brightness temperature threshold, and it is certain that the VLBI component in \tar\ is of non-thermal origin.

VLBI imaging observations of nearby Seyfert galaxies and radio-quiet AGN with the radio flux density $>1$ mJy show that the majority of these sources are dominated by an unresolved radio core, and a few exhibit a compact symmetric object (CSO) morphology on scales of a few parsec \citep{1998MNRAS.299..165B,2005ApJ...621..123U}. The fraction of the compact cores we detected has a tendency to increase with the total flux density \citep{1998MNRAS.299..165B}. The brightness temperatures of compact cores exceed $10^8$ K \citep[e.g.][]{1996ApJ...468L..91B,2005ApJ...621..123U}. The brightness temperature of \tar\ is consistent with the temperatures of the typical Seyfert I galaxies and radio-quiet AGN, suggesting that they may have the same physical origin. Therefore, a natural explanation of the compact VLBI component detected in \tar\ is the radio core of the AGN.

In addition to the possibility of the radio core (an optically thick weak jet), there may be other origins of compact radio emission in the galactic nucleus.
Ultrafast outflows have been observed in optical emission lines and X-ray continuum in \tar, which may collide with the interstellar medium to produce shocks. In addition, flares produced during disc crossing or pericentre passage can also produce shocks in the corona \citep{2022arXiv220111633J}. All these shocks may produce synchrotron radiation in the radio band.

Although the high brightness temperature excludes optically thick thermal radiation, which cannot exceed the temperature of the ionised gas, the brightness temperature alone does not fully rule out the presence of compact optically thin thermal emission originated from ionised disc  winds \citep{2007ApJ...668L.103B}. 
The spectral index calculated from the VLBA observations taken one day apart is $\alpha^{4.9}_{1.6} = 0.04 \pm 0.18$  (the spectral index is defined as $S_\nu \propto \nu^\alpha$). The spectral index obtained from the EVN 1.66 GHz and VLBA 4.9 GHz observations three days apart is $\alpha^{4.9}_{1.7} = -0.21 \pm 0.14$. Although the flux densities from the VLBA data may be slightly lower than the actual values because the pointing centre deviated by $\sim1\arcsec$ from the source position (see Section \ref{sec:obs}), both spectral indices consistently show a flat spectrum, so that the possibility of an optically thin steep-spectrum disc wind becomes implausible. 
Moreover, if wide-angle outflow structures were present, they would be resolved by the high-resolution VLBI into a number of clumpy features \citep[see an exmaple in the superwind quasar PDS~456;][]{2021MNRAS.500.2620Y}. However, this is obviously incompatible with the \tar\ VLBI images, which show a single compact component. These pieces of evidence do not favour the wide-angle outflow or wind interpretation.

A magnetically heated corona above the accretion disc can also produce radio emission, which has a very compact structure becaues the radiation region lies very close to the central engine, and it has a flat spectrum due to synchrotron self-absorption \citep{2019NatAs...3..387P}. The radio source produced by the accretion disc corona resembles the radio core in many observational characteristics and is considered to be one of the origins of radio emission from radio-quiet AGNs. One criterion for identifying coronal activity is the radio to X-ray luminosity ratio, which satisfies the G\"{u}del-Benz relation, that is, $L_R/L_X \sim 10^{-5}$ \citep{2008MNRAS.390..847L}. We examined the radio and X-ray data of \tar\ on 2022 March 2. 
The radio luminosity of the VLBI component is 
$3.9\times10^{21}$ W Hz$^{-1}$ at 4.9 GHz,
consistent with the radio luminosities of nearby Seyfert galaxies, which are between $10^{20} - 10^{23}$ W Hz$^{-1}$ \citep{1989ApJ...343..659U,2021MNRAS.500.4749B}. 
The X-ray luminosity in the 2--10 keV band is 
$L_X = 3.5\times10^{43}$ erg s$^{-1}$ on 2022 March 2.
This gives $L_R/L_X = 5.5\times10^{-6}$, a factor of 0.55 of the G\"{u}del-Benz relation \citep{2008MNRAS.390..847L}, suggesting that both radio and X-ray emission are likely to be from coronal activity.

Compact radio sources may also be young radio supernovae (RSNe), such as the RSNe observed in the southern nuclei of the dual AGN system NGC~6240 \citep{2011AJ....142...17H}. VLBI observations from the so-called supernova factories Arp~220 and Arp~299 show that the vast majority of RSNe have luminosities below $10^{20}$ W Hz$^{-1}$, and only a very few are brighter than $10^{21}$ W Hz$^{-1}$ \citep{2009AJ....138.1529U,2019A&A...623A.173V}. The  radio luminosity of RSNe is clearly not comparable with that of the \tar\ VLBI component. Moreover, RSNe usually have a steep spectrum ($-0.7< \alpha <-0.5$), which is not consistent with the flat spectrum of \tar. This further rules out the possibility of a RSNe.
To summarise, a consistent conclusion is obtained from the radio morphology, size, brightness temperature, radio spectral index, and radio-X-ray luminosity ratio that the compact component detected in the VLBI images is either a compact jet or associated with the corona.

\subsection{Understanding the radio structure}

In the previous section, we discussed the physical nature of the compact VLBI component. In this section, we study the radio emission on larger scales. 

In parallel with the VLBI observations, we also carried out radio observations on the arcsecond scale to obtain the radio emission properties of the whole galaxy, including using the ATCA, JVLA, and uGMRT.
The ATCA observations are the closest in time to the VLBI observations. The preliminary observational results from the ATCA have been reported in ATel\#15267 \citep[][]{2022ATel15267....1A}. The observations were carried out on 
2022 February 28
UT 15-21, and \tar\ is detected at all three frequencies of 2.1, 5.5, and 7.5 GHz. The flux density at 2.1 GHz is 0.76$\pm$0.05 mJy. 
Recently, Bruni et al. (ATel\#15306) reported an observation made with the 
e-MERLIN
on 2022 March 28, detecting a compact source with a flux density of 0.73$\pm$0.09 mJy at 1.5 GHz with a deconvolved size of 85 mas $\times$ 34 mas. 
Despite the difference in resolution, the measured flux density by e-MERLIN with sub-arcsec resolution is very consistent with those by ATCA and uGMRT around 1.5 GHz with arcsec resolutions, suggesting that there is only very small room for a $>100$ mas scale extended emission contributed by galactic-scale star formation. This also indicates that the radio emission has remained stable over about one month after the latest optical and X-ray flares in the end of January \citep{2022arXiv220111633J}.

This e-MERLIN flux density is about 2.3 times the VLBA and EVN flux densities at 1.6 GHz, suggesting that
about 40\% of the radio emission of \tar\ comes from  the compact feature within $\sim$0.8 pc of the nuclear region. \textit{\textup{The question now is where the remaining 60\% of the radio flux density originates.} }
We searched a larger region of 200$\times$200 mas$^2$ in the EVN and VLBA images centred on the location of the VLBI peak and found no other compact components above $5\sigma$, except for the radio core. 
Therefore, the remaining radio emission is probably from a scale between 12 (the lowest resolution of the VLBA at L band) and 120 mas  (the highest resolution of the e-MERLIN). Extended jets and extended NLR winds are both possible on this size scale.
This is similar to multi-scale radio observations of other Seyfert galaxies, in which a flat-spectrum core  contributes a large fraction of the total flux density at centimetre wavelengths \citep{1989ApJ...343..659U}. The fraction of the radio core flux density seems to be correlated with the total flux density and observation frequency: the higher the total flux, the larger this fraction. At 4.9 GHz, a somewhat higher percentage of  extended radio emission is resolved.

An extended jet is a natural explanation of the missing flux in VLBI images. First of all, it is certain that \tar\ lacks an extended jet on $\gtrsim 100$ pc scales. Compared to the highly relativistic jets in blazars, jets from radio-quiet AGNs have significantly lower kinetic power, a less collimated structure (or the jet direction varies), and they are likely to be intermittent (see Mrk~231 for a similar situation;  \citealp{2009ApJ...706..851R,2021MNRAS.504.3823W}). On the other hand, \tar\ may have a stratified or structured jet, and the entrainment of gas from the NLR clouds or torus into the outer sheath of the jet may result in hydrodynamic instability and lead to the fragmentation of the jet flow \citep[see an example of the disrupted jet in 3C~48;][]{1991Natur.352..313W,2010MNRAS.402...87A}. These factors hamper the 
expansion 
of the low-power jet to large scales, 
confining the jet flow
within a cocoon \citep[e.g.][]{2007MNRAS.377..731M}. On the one hand, mildly relativistic jets are continuously generated in the nucleus, and the brightness of the jet knot rapidly declines through adiabatic expansion as it moves outwards. On the other hand, shocks are formed at the interface with the cocoon in the direction of the jet, producing the lobe structure commonly seen in CSO galaxies \citep{2021A&ARv..29....3O}. As a result, the radio structure consists of a central compact core, lobes at a few tens of parsec, and intervening invisible faint jets. e-MERLIN can detect all these emission features, but the extended lobes are completely resolved in the VLBI images. This type of structure is commonly seen in many Seyfert galaxies.

Alternatively, the nuclear outflows can produce the extended radio emission.
The Sloan Digital Sky Survey spectrum of \tar\ exhibits blueshifted broad $H_\alpha$ emission line feature with a velocity offset of 2400 km s$^{-1}$ (see Figure 2 of \citealt{2022arXiv220111633J}), which in the SMBHB scenario is caused by the Doppler shift of the $H_\alpha$ line from BLR clouds as the secondary SMBH orbits the primary SMBH. 
The BLR clouds might be tidally disturbed and scattered by the primary black hole, resulting in additional high-velocity unbound components that may become ionised outflows driven by radiation pressure of the disc. 
High-velocity inflow and outflow components are also found in the X-ray continuum.
Powerful high-speed AGN winds or outflows \citep[e.g.][]{2014MNRAS.442..784Z,2015MNRAS.447.3612N,2015Sci...347..860N,2018ApJ...854L...8R} can generate shocks that may affect scales from a few parsec to 100 
parsec
and produce synchrotron radiation in radio. Shocks created from the interactions between parsec-scale winds or outflows with a dense external medium can be distributed over a large opening angle. For instance, the E and F clumps discovered in the Seyfert galaxy NGC~3079, which deviate from the jet direction, are considered to be rapidly cooling remnants of interactions of the outflow with the interstellar medium \citep{2005ApJ...618..618K}.  Studying the energetics of the outflow at multiple bands can help confirm or rule out whether it can produce the observed radio luminosity.  

\subsection{Accretion state and jet-disc coupling}

From the trajectory models that match the optical  light curves, the evolution of the binary orbit predicts a merger time of about 3 years \citep{2022arXiv220111633J}. 
Ongoing X-ray monitoring shows the latest flare at the end of 
2022 January,
after which the X-ray brightness decreases. For the dates of the radio VLBI observations, 2022 February 27 and 2022 March 2, the X-ray luminosities are at a local trough, with X-ray luminosities of $L_{\rm X} (\rm 2-10\, keV) = 5.9\times10^{43}$ erg s$^{-1}$ and $3.5\times10^{43}$ erg s$^{-1}$, respectively. 

The classical radio-loudness parameter $R$ is defined as the ratio of two monochromatic fluxes $R = f(6 cm)/f(4400 \AA)$ \citep{1989AJ.....98.1195K}, where $R<10$ for radio-quiet AGN. The monochromatic flux density at rest-frame $4400\AA$  derived from optical spectra taken on 
2022 February 28 
is $f(4400\AA)=0.24$~mJy, yielding a radio-loudness parameter of $R \sim 1$.
By definition, \tar\ is a radio-quiet AGN. Alternatively, the  monochromatic luminosity at the 6 cm radio band and X-ray luminosity can be used to define the radio loudness \citep{2003ApJ...583..145T}, $R_X=\nu L_\nu {\rm (6 \,cm)}/L_X (2-10 \, {\rm keV})$. This gives a radio loudness of $R_X \sim 10^{-5}$, again confirming the radio-quiet nature. 

Previous studies have found a correlation between black hole accretion and jets, specifically, between the radio core luminosity at 5 GHz ($L_{R}$), the X-ray luminosity ($L_X$) at 2--10 keV, and the black hole mass ($M_{\rm BH}$), that is, the black hole fundamental plane (FP) relation \citep[e.g.][]{2003MNRAS.345.1057M,2004A&A...414..895F}, 
\begin{equation}
    \log L_{\rm R} = (0.60^{+0.11}_{-0.11}) \log L_{\rm X} + 
        (0.78^{+0.11}_{-0.09})\log M_{\rm BH} + 7.33^{+4.05}_{-4.07} \label{eq1}
.\end{equation}
The extended radio emission observed in radio-quiet AGN is associated with a relic from earlier radio activity, and it is not a contemporaneous event with the current X-ray activity. Therefore, we 
used the radio luminosity $L_R$ obtained from the VLBI measurements, which is much closer to the recent X-ray activity.
Substituting the observed X-ray luminosity and $M\sim 10^8 M_\odot$ \citep{2022arXiv220111633J} into the black hole FP relation, we find that the inferred radio flux density is more than one order of magnitude higher than the observed value. 
This discrepancy may be related to the enhanced X-ray luminosity during the coalescence phase that is used in the calculation. If we substitute the observed radio luminosity into Eq.~\ref{eq1}, we obtain an X-ray luminosity that is two orders of magnitude lower than the observed value, which could be the quiescent-state X-ray luminosity and needs to be confirmed by future X-ray observations.
We note that the data points presented in
\citet{2003MNRAS.345.1057M} for fitting the FP relation can vary by more than one order of magnitude, so that the physical implication that can be drawn from the mismatch obtained here is unclear.

On the other hand, the black hole FP relation is affected by various AGN sample properties (e.g. radio spectral index or beaming effect) and different AGN accretion states (high or low states).
Under the assumption of a scale-invariant jet, the relation is based on the existence of a coupling between the radiatively inefficient accretion flow and the jet, independent of the jet model, and on the assumption that the correlation coefficient depends only on the spectral index of the synchrotron radiation and the accretion mode. 
Whereas Seyfert galaxies are typically thought to contain radiatively efficient discs, the FP of Seyfert galaxies may not be the same as that obtained from low-luminosity AGN \citep{2019ApJ...871...80G}. If two black holes are present, the sources of X-ray and radio emission remain rather uncertain for the particular AGN studied
here. Future multi-band observations may help understand the origin and location of the emitters.

For X-ray binaries, observations indicate that
the radiatively inefficient accretion flow can produce a weak steady jet when the accretion rate or X-ray emission is low; 
when the accretion state becomes higher to allow stronger X-ray emission, the jet production is suppressed. 
When the binary black holes merge, the accretion flow is likely significantly disturbed and later on settles to a new stationary state. A new transient jet may be produced at that time. Our current VLBI observations provide the necessary information for understanding the entire pre-merger 
radio emission of the SMBHB.

\section{Conclusions} \label{sec:conc}

We have conducted observations of \tar\ using the European VLBI Network and the Very Long Baseline Array and obtained high-resolution images of this source for the first time.
The main results and conclusions of this paper are summarised below.
   \begin{enumerate}
      \item A highly compact radio component with a brightness temperature higher than $10^8$~K and a flat radio spectrum is consistently detected in all three VLBI images. Based on its radio morphology, size, brightness temperature, and radio spectrum, we identify the VLBI component as an optically thick jet (i.e. a radio core) and/or a corona.
      \item The precise position of the nucleus with an accuracy of 0.1 mas is determined, paving the way for future VLBI monitoring.
      \item Approximately 40\% of the total radio flux comes from the compact component revealed in the VLBI images. The remaining 60\% of the flux density may come from extended radio lobes that extend to a few tens of parsecs from the nucleus, from the resolved outer layer of the structured jet, or from shocks generated by high-speed ionised outflows in the NLR.
      \item A mildly relativistic jet such as that in \tar\ may be confined within the NLR of the host galaxy. Similar observations have been found in other radio-quiet AGNs and radio-intermediate AGNs (e.g. Mrk~231). In contrast, powerful Fanaroff-Riley I type jets, even if the jets have experienced strong interactions with the interstellar medium and lost a fraction of their mechanical energy (e.g. the distorted jet in 3C~48), still have high enough momentum to break through the interstellar medium to reach beyond 1 kpc.
      \item 
      Based on the black hole FP relation, the X-ray luminosity estimated from the VLBI radio luminosity and black hole mass is two orders of magnitude lower than the observed values, and it may be associated with the radiatively inefficient accretion flow and not be identical to the currently active X-ray emission source(s). As the SMBHB coalescence is completed, a stable accretion disc  will form and the accretion state may change significantly, resulting in a possible jet after the coalescence.
   \end{enumerate}
   
The combined multi-band observations on future disc crossing events will be able to give a more stringent constraint on the SMBHB evolutionary model. In the coming years, continued radio spectrum and VLBI imaging observations will provide key observational information to track the last moments before the binary SMBH coalescence.

\begin{acknowledgements}
We thank the anonymous referee for the constructive comments and suggestions, the inclusion of which has improved the content and presentation of our work.
This work is supported by National Key R\&D Programme of China (grant number 2018YFA0404603) and NSFC (1204130, 11833007, 12192221, 12022303). T.A. thanks for the grant support by the Youth Innovation Promotion Association of CAS. Y.Z. is sponsored by Shanghai Sailing Program under grant number 22YF1456100. H.Y. and Z.P. acknowledge supports by the Natural Sciences and Engineering Research Council of Canada and in part by Perimeter Institute for Theoretical Physics. Research at Perimeter Institute is supported in part by the Government of Canada through the Department of Innovation, Science and Economic Development Canada and by the Province of Ontario through the Ministry of Colleges and Universities.
We thank the TACs and schedulers of the EVN and NRAO for approving the ToO/DDT observations and scheduling the observations quickly. We thank the staff at the NRAO and EVN telescopes and correlators for observing and quickly correlating the data.
The European VLBI Network (EVN) is a joint facility of independent European, African, Asian, and North American radio astronomy institutes. Scientific results from data presented in this publication are derived from the following EVN project code(s): RS005. 
e-VLBI research infrastructure in Europe is supported by the European Union, Seventh Framework Programme (FP7/2007-2013) under grant agreement number RI-261525 NEXPReS. e-MERLIN is a National Facility operated by the University of Manchester at Jodrell Bank Observatory on behalf of STFC. 
The National Radio Astronomy Observatory is a facility of the National Science Foundation operated under cooperative agreement by Associated Universities, Inc. Scientific results from data presented in this publication are derived from the following VLBA project codes: BA154.
The VLBI data processing made use of the compute resource of the China SKA Regional center prototype, funded by the Ministry of Science and Technology of China and the Chinese Academy of Sciences.
\end{acknowledgements}

%
  \bibliographystyle{aa} 
  \bibliography{refs.bib} 
%

\clearpage
\begin{table*}
\centering
\small
\caption{Observation setup of the VLBA sessions on J1430+2303. \label{tab:obs}}
\begin{tabular}{ccccccc} \hline\hline
Session & Date  & $\nu_{obs}$ & Time$_{obs}$ & Bandwidth & Antennas & PR Calibrator\\
        & Y-M-D & (GHz)& (hour) & (MHz) & \\ \hline
RS005  & 2022-02-27 &  1.66   & 5.5  &  256   MHz  & Ef Mc O8 Tr Wb Nt T6 Ur Hh Sv Zc Bd Ir Jb & J1427$+$2348\\
BA154A & 2022-03-02     &  1.57   & 4.5  &  512 MHz  & SC,FD,LA,BR,HN,NL,PT,KP,OV,MK & J1427$+$2348\\
BA154B & 2022-03-03     &  4.90   & 4.5  &  512 MHz  & SC,FD,LA,BR,HN,NL,PT,KP,OV,MK & J1427$+$2348\\ \hline
\end{tabular} \\
The EVN telescopes participating in RS005 are Jodrell Bank Mk2 (Jb, United Kingdom), Westerbork (Wb, The Netherlands), Effelsberg (Ef, Germany), Medicina (Mc, Italy), Noto (Nt, Italy), Onsala 25-m (O8, Sweden), Toru\'{n} (Tr, Poland), Irbene 32-m (Ir, Latvia), Svetloe (Sv, Russia), Zelenchukskaya (Zc, Russia), Badary (Bd, Russia), Tianma (T6, China), Hartebeesthoek (Hh, South Africa). The VLBA telescopes participating in these observations are BR (Brewster,), FD (Fort Davis), HN (Hancock), KP (Kitt Peak), LA (Los Alamos), MK (Mauna Kea), NL (North Liberty), OV (Owens Valley), PT (Pie Town) and SC (Saint Croix).
\end{table*}

\begin{table*}
\centering
\small
\caption{Phase-reference imaging results of J1430+2303. \label{tab:img}}
    \begin{tabular}{cccccccc} \hline\hline 
Freq.   & S$_{peak}$    & S$_{tot}$ & Position (J2000) & Beam & r.m.s.  & D$_{size}$   & T$_{b}$ \\ 
(GHz) & (\mJyb) & (mJy) & & (mas $\times$ mas, \degr) & (\uJyb) & (mas) & ($10^7$~K) \\ \hline 
    1.57 (VLBA L band) & 0.22$\pm$0.02 & 0.23$\pm$0.03 & 14:30:16.04096, $+$23:03:44.5387 & 11.7$\times$5.1, 1.6\degr  & 18  & $<2.2$ & $>2.9$  \\
    1.66 (EVN L band) & 0.30$\pm$0.02 & 0.30$\pm$0.03 & 14:30:16.04095, $+$23:03:44.5391 & 7.6$\times$2.9, 13.5\degr  & 15  & $<1.0$ & $>14.0$ \\
    4.90 (VLBA C band) & 0.24$\pm$0.02 & 0.24$\pm$0.02 & 14:30:16.04093, $+$23:03:44.5390 & 3.9$\times$1.7, $-5.4\degr$& 8  & $<0.5$ & $>5.9$ \\ \hline 
        \end{tabular} \\ 
\end{table*}

\end{document}